\documentstyle[prl,aps,epsf,12pt]{revtex}
\newcommand{\beq}{\begin{equation}}
\newcommand{\eeq}{\end{equation}}
\newcommand{\bea}{\begin{eqnarray}}
\newcommand{\eea}{\end{eqnarray}}

\begin{document}

\baselineskip=24pt

\title{Proton Single Particle Energy Shifts due to Coulomb Correlations }

\author{Aurel Bulgac $^{ 1 }$ and Vasily R. Shaginyan $^{1,2}$}

\address{ $^{ 1 }$ Department of Physics, University of Washington,
Seattle, WA 98195, USA}

\address{ $^{ 2 }$ Petersburg Nuclear Physics Institute, 188350 Gatchina,
RUSSIA}

\maketitle

\begin{abstract}

A theoretically consistent approach to the calculation of the Coulomb
correlation corrections to the single--particle energies is
presented. New contributions to the single--particle energies
previously overlooked in the literature are now identified and taken
into account.  We show that the interplay between the Coulomb
interaction and the strong interaction, which is enhanced in the
nuclear surface, leads to an upward shift of the proton
single--particle levels. This shift affects the position of the
calculated proton drip line, a shift towards decreasing $Z$. We
describe briefly a similar mechanism which is at work for neutron
levels.  The same mechanism is responsible for significant corrections
to the mass difference of the mirror nuclei (Nolen--Schiffer anomaly)
and to the effective proton mass.

\end{abstract}



\vspace{0.3cm}

The main part of the Coulomb energy in nuclei is given by the Hartree
contribution and to a reasonable  accuracy this can be computed as the
energy of a uniformly charged sphere and is thus proportional to
$Z^2/A^{1/3}$. There are a number of corrections, some of them rather
subtle, arising from the interplay between the Coulomb and nuclear
forces. The Nolen--Schiffer anomaly in the binding energy of
mirror nuclei \cite{n} is a case in point. In Ref. \cite{av} we have
shown that a specific many--body mechanism (outlined concisely below)
leads to an enhancement of the Coulomb energy in the nuclear surface
region and in this way one can account for a major part of this
anomaly. This effect results in a systematic contribution to the
nuclear binding energy, which scales as $\propto Z^{2/3}$. We are
going to demonstrate that the mechanism  should be taken into
account when calculating a number of nuclear properties. In this
Letter we study the role of this new many--body effect  on the
single--particle proton energy levels, on the location of the proton
drip line and on the proton effective mass.

We shall operate within the density functional theory
\cite{dft,kks,fa,fttz}. The ground state energy of a nucleus $E$ is
given by a sum of two functionals (in the absence of
pairing correlations):
\beq
E=F_0[\rho_\pi({\bf r}),\rho_\nu ({\bf r})]\,
+\,F_{Coul}[\rho_\pi({\bf r}), \rho_\nu({\bf r})],\eeq
where the symmetric part $F_0[\rho_\pi({\bf r}),\rho_\nu({\bf r})]=
F_0[\rho_\nu({\bf r}),\rho_\pi({\bf r})]$ is due to the (strong)
isospin conserving nuclear forces, while $F_{Coul}$ is due to the
(weak) Coulomb interaction. We shall neglect in our analysis several
easy to include terms: the trivial contribution in the kinetic energy,
arising from proton--neutron mass difference, the contribution of the
Charge Symmetry Breaking (CSB) forces \cite{csb}. For the sake of
simplicity of the presentation, we shall not display explicitly the
spin degrees of freedom and the contribution arising from the
spin--orbit interaction, even though we have included them in the
actual calculations. The proton and neutron densities are defined as
usual
\bea
\rho_\pi({\bf r})&=&\sum_l n_{\pi\;l}|\phi_{\pi\;l}({\bf r})|^2,\\
\rho_\nu({\bf r})&=&\sum_l n_{\nu\;l}|\phi_{\nu\;l}
({\bf r})|^2, \eea
where $n_{\pi\;l}, \phi_{\pi\;l}$ and $n_{\nu\;l}, \phi_{\nu\;l}$ are
proton and neutron quasiparticle occupation numbers and
single--particle wave functions respectively. The well known Skyrme
functional, see e.g. Refs.  \cite{rs,bab}, can be considered as one
possible realization of $F_0$. The Coulomb energy functional is given
by:
\bea
F_{Coul}&=&e^2\int d{\bf r}_1d{\bf r}_2
\frac{\rho_{\pi\pi} ({\bf r}_1,{\bf r}_2)}{|{\bf r}_1-{\bf r}_2|}
= F_{Coul}^{Hartree}+
F_{Coul}^{Fock}+
F_{Coul}^{corr}
\label{eq:Cexact},\\
F_{Coul}^{Hartree}
&=& \frac{e^2}{2}\int d{\bf r}_1d{\bf r}_2
\frac{\rho_\pi({\bf r}_1)\rho_\pi({\bf r}_2)}{|{\bf r}_1-{\bf r}_2|}
\label{eq:CHartree},\\
F_{Coul}^{Fock}
&=& -\frac{e^2}{2}\int  \frac{d{\bf r}_1d{\bf r}_2 d\omega}{2\pi}
\frac{ \chi^0_\pi({\bf r}_1,{\bf r}_2,i\omega)
+2\pi\rho_\pi ({\bf r}_1)\delta({\bf r}_1-
{\bf r}_2)\delta(\omega)}{|{\bf r}_1-{\bf r}_2|}
\label{eq:CFock},\\
F_{Coul}^{corr} &=&
-\frac{e^2}{2}\int \frac{d{\bf r}_1d{\bf r}_2 d\omega}{2\pi}
\frac{\chi_{\pi\pi}({\bf r}_1,{\bf r}_2,i\omega)-
\chi^0_\pi({\bf r}_1,{\bf r}_2,i\omega)}{|{\bf r}_1-{\bf r}_2|},
\label{eq:Ccorr}
\eea
with $\rho_{\pi\pi}({\bf r}_1,{\bf r}_2)$ being the exact two--proton
density distribution function.  $\chi_{\pi\pi}({\bf r}_1,{\bf
r}_2,i\omega)$ and $\chi^0_\pi({\bf r}_1,{\bf r}_2,i\omega)$ are the
full and free proton linear response functions respectively, evaluated
at the imaginary frequency $i\omega$.  Eqs. (\ref{eq:CHartree}) and
(\ref{eq:CFock}) represent the Hartree and Fock contributions to the
Coulomb energy. The exchange term is written here in a somewhat
unusual way \cite{av}, through the linear response function of the
noninteracting protons, $\chi^0_\pi$, since upon integrating along the
real axis of the complex $\omega$ plane one has
\beq
\int_0^{\infty}\frac{d\omega}{\pi}{\rm{Im}}
\chi^0_\pi({\bf r}_1,{\bf r}_2,\omega)=
[\rho_\pi({\bf r}_1,{\bf r}_2)\rho_\pi({\bf r}_2,{\bf
r}_1)-\delta( {\bf r}_1-{\bf r}_2)\rho_\pi({\bf r}_1)],
\eeq
where $\rho_\pi ({\bf r}_1,{\bf r}_2)$ is the proton single--particle
density matrix \cite{erratum}.  In
Eqs. (\ref{eq:CFock},\ref{eq:Ccorr}) we have performed a Wick rotation
in order to evaluate the integral along the imaginary axis.  Upon
taking into account the (strong) residual interaction, the linear free
response function $\chi^0_\pi$ should be replaced with the full
response function $\chi_{\pi\pi}$ and one thus readily obtains the
expression for the Coulomb correlation energy (\ref{eq:Ccorr}). By
considering all three contributions
(\ref{eq:CHartree},\ref{eq:CFock},\ref{eq:Ccorr}) we therefore account
for all diagrams in first order in $e^2$, in the (weak) Coulomb
interaction. The (strong) nuclear interaction is treated to all
orders. We are going to concentrate on a study of the contribution to
the single particle energies and the effective mass due to
Eq. (\ref{eq:Ccorr}). In passing we remark that there are no
ambiguities with double counting, rearrangement energy,
core--polarization correction and so forth, see Refs. \cite{shl} for a
discussion of this issues, within the density functional formalism.

By expressing $F_{Coul}^{corr}$ through the linear response function
we can easily calculate a number of functional derivatives needed
below. At the same time one can clearly disentangle the contribution
of various modes to this part of the energy density functional as well
\cite{kks}. In Ref. \cite{av} we have shown that the main contribution
to $F_{Coul}^{corr}$ comes from the surface collective isoscalar
excitations. In the isoscalar channel the particle--hole residual
interaction has a strong density dependence, changing from a
relatively weak one inside nuclei to a strong attractive one in the
surface region \cite{migdal}. Because of the attractive character of
the isoscalar residual particle--hole interaction, the nuclear surface
(where the matter density is low) is very close to instability
\cite{av}. Low density homogeneous nuclear matter is manifestly
unstable and this shows itself in the fact that the linear response
function has a pole for imaginary values of $\omega$ \cite{av}. In
finite nuclei and semi--infinite nuclear matter this singularity is
smoothed out and becomes a prominent peak of the response function at
low frequencies in the surface region. This proximity of the nuclear
surface to instability is the reason why the contribution of the
(weak) Coulomb exchange interaction is strongly enhanced and the
correlation Coulomb ``correction'' to the mass formula has as a result
a predominantly surface character, i.e. $F_{Coul}^{corr}\propto
Z^{2/3}$.

Let us turn now to the calculation of the proton single--particle
energy corrections due to the presence of $F_{Coul}^{corr}$ in the
energy functional. Using Landau's variational equation \cite{lan}
\beq
\varepsilon_{\pi\;l} = \frac{\delta E}{\delta n_{\pi\;l}},
\eeq
one obtains for the proton single--particle energy shift
$\Delta \varepsilon_{\pi\;l}$ the following expression
\beq
\Delta\varepsilon_{\pi\;l}
=-\frac{e^2}{2}\int
\frac{d{\bf r}_1d{\bf r}_2 d\omega}{2\pi |{\bf r}_1-{\bf r}_2| }
\frac{\delta 
[\chi_{\pi\pi}({\bf r}_1,{\bf r}_2,i\omega)-
\chi^0_\pi({\bf r}_1,{\bf r}_2,i\omega)]}{\delta n_{\pi\;l}} .
\label{eq:spe}
\eeq
The variational derivative $\delta\chi^0_\pi/\delta n_{\pi\;l}$
has the simple functional form,
\beq
\frac{\delta\chi^0_\pi({\bf r}_1,{\bf r}_2,i\omega)}
{\delta n_{\pi\;\lambda_0}}
=\left[G_\pi({\bf r}_1,{\bf r}_2,i\omega
+\varepsilon_{\pi\;\lambda_0})
+(\omega\rightarrow -\omega ) \right]
\phi^*_{\pi\;\lambda_0}({\bf r}_1)
\phi_{\pi\;\lambda_0}({\bf r}_2),
\eeq
with $G_\pi({\bf r}_1,{\bf r}_2,\omega)$ being the single--particle
proton propagator in the selfconsistent nuclear potential. The linear
response function $\chi _{\pi\pi}$ is obtained by solving the usual
matrix functional equation (Landau zero sound or RPA)
\beq
\chi _{ij}=\chi^0_i\delta_{ij}
+\sum_{k=\pi,\nu} \chi^0_iR_{ik}\chi_{kj},
\eeq
where $R_{ik}$ is the irreducible (strong) particle--hole interaction
and $i,j$ and $k$ stand for the isospin variables, $\pi $ and $\nu$
for protons and neutrons respectively.  From this equation one derives
the following equation for $\delta\chi_{\pi\pi}/\delta n_{\pi\;l}$
\beq
\frac{\delta\chi_{ij}}{\delta n_{\pi\;l}}=
\frac{\delta\chi^0_i}{\delta n_{\pi\;l}} \delta_{ij}
+\sum_{k=\pi,\nu}\left[
\frac{\delta\chi^0_m}{\delta n_{\pi\;l}} R_{ik}\chi_{kj}
+\chi^0_{i}   \frac{\delta R_{ik}}{\delta n_{\pi\;l}}\chi_{kj}
+\chi^0_{i} R_{ik} \frac{\delta\chi_{kj}}{\delta n_{\pi\;l}} \right].
\eeq
The operator solution can be easily obtained and one thus can show
that the relevant quantity in the integrand of Eq. (\ref{eq:spe}) has
the following structure
\bea
& & \frac{\delta\chi}{\delta n_{\pi\;l}} 
-\frac{\delta\chi _0}{\delta n_{\pi\;l}} 
=
\frac{1}{1-\chi _0 R}
\frac{\delta\chi^0}{\delta n_{\pi\;l}}
\frac{1}{1-R\chi_0} -
\frac{\delta\chi _0}{\delta n_{\pi\;l}} 
+\chi \frac{\delta R}{\delta n_{\pi\;l}}\chi   \label{eq:dchi} \\
&=& \frac{\delta\chi _0}{\delta n_{\pi\;l}} R\chi +
\chi R \frac{\delta\chi _0}{\delta n_{\pi\;l}} 
+\chi R \frac{\delta\chi _0}{\delta n_{\pi\;l}} R \chi
+\chi \frac{\delta R}{\delta n_{\pi\;l}}\chi \nonumber
\eea
where for the sake of clarity we have suppressed the isospin, spin
and spatial coordinates and the corresponding summations and integrations.

\begin{figure}[h,t,b]
\begin{center}
\epsfxsize=9.0cm
\centerline{\epsffile{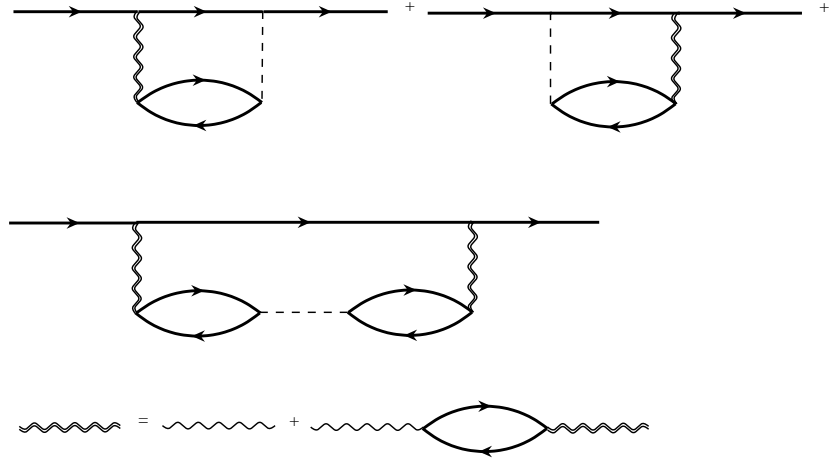}}
\end{center}
\caption{The diagrammatic representation of corrections to the
single--particle nucleon propagator, corresponding to the Coulomb
correlation energy discussed in this Letter.
The solid oriented lines represent the nucleon propagators, the dashed
lines the Coulomb interaction, the thin wavy line the strong
nuclear interaction $R_{ik}({\bf r}_1,{\bf r}_2)$ and the double
wavy lines stand  for the ``screened'' strong nuclear interaction, 
obtained by solving the diagrammatic equation shown in the third row.
}
\label{fig1}
\end{figure}

In Fig. 1 we show the diagrammatic representation of the corrections
to the single--particle nucleon propagator, corresponding to the
addition of the Coulomb correlation energy to the nuclear density
functional given by Eq. (7). Strictly speaking, we have omitted in
this diagrammatic representation the contribution arising from the
last term in Eq. (\ref{eq:dchi}), which depends on the functional
derivative of the irreducible particle--hole interaction $\delta
R/\delta n_{\pi\;l}$. We shall discuss in more detail the relevance of
such a correction in a future publication. To the best of our
knowledge this correction was not explicitly discussed in literature
in this context. A superficial analysis would suggest that this type
of correction is perhaps small. The two diagrams shown in the
upper row were considered before by other authors \cite{shl}, however
the third diagram, shown in the middle row, was, surprisingly,
overlooked. All three diagrams are of the same order in the
electromagnetic coupling constant $e^2$. Only when considering all
three diagrams does the correction to the single--particle energy
satisfy the natural Landau's variational equation (9). Besides being
theoretically consistent, the present approach has a new physical
feature as well: there are corrections to the single--particle
neutron properties too. While for protons all three diagrams
have to be considered, for neutrons there is a Coulomb correlation
correction to the single--particle energies arising from the third
diagram alone.

We do not consider in this Letter the so called AKW diagrams
\cite{akw}, which are automatically taken into account in
selfconsistent calculations of nuclear masses \cite{bab}.  Auerbach
\cite{na} has shown that the corrections corresponding to these
diagrams are too small to account for the Nolen--Shiffer anomaly.

In Ref. \cite{av} we have used a simple separable model for the
residual interaction $R_{ik}$
\beq
R_{ik}({\bf r}_1,{\bf r}_2) = \lambda\frac{d V_i(r_1)}{dr}
\frac{d V_k(r_2)}{dr}\delta(\Omega_1-\Omega_2),
\eeq
where $ V_i(r)$ is the proton/neutron single--particle potential in
a spherical nucleus and $\delta(\Omega_1-\Omega_2)$ is a Dirac
function in angle variables. $\lambda $ is chosen so that the dipole
linear response has a pole at $\omega =0$, corresponding to the
spurious mode.  This type of residual interaction has been studied
extensively \cite{rs,bm,eb} and it leads to a satisfactory description
of nuclear collective modes. We have used the same nonselfconsistent
approach described in detail in Ref. \cite{av} in order to estimate
the magnitude of the correction $\Delta \varepsilon_{\pi\;l}$. For
various single-particle proton levels around the Fermi level and the
proton threshold, the calculated shifts $\Delta \varepsilon_{\pi\;l}$
are in the interval $0.1-0.3$ MeV in light ($A=16$) and medium
($A=40-48$) nuclei. In performing these calculations in each nucleus
we have included collective modes with multipolarities up to $\approx
2A^{1/3}$ and energies up to $\approx 225 $ MeV, see Ref. \cite{av}.
Both in multipolarities and in energy convergence was attained in our
calculations. The values of $\Delta \varepsilon_{\pi\;l}$ thus
obtained are of the same magnitude as the Nolen--Schiffer anomaly.

It is instructive to cross check these results using an independent
approach. A new type of nuclear density functionals has been recently
introduced in Refs. \cite{fa,fttz}, see also Ref. \cite{bab}.  The
main reason for seeking new functionals is to obtain a significantly
more accurate reproduction of the nuclear properties (bindings
energies and matter distribution of finite nuclei and infinite neutron
and symmetric nuclear matter properties over a wide range of densities
simultaneously) than one can achieve with the plethora of existing
density functionals. A key ingredient was the introduction of the
Coulomb correlation energy contribution, along the lines suggested
earlier by us in Refs. \cite{av}. However, while we have presented
arguments for a significant surface contribution into the nuclear
Coulomb correlation energy, Fayans \cite{fa} has chosen to parametrize
the Coulomb correlation energy as a volume term:

\beq
F_{Coul}^{corr}[\rho_\pi({\bf r}),\rho_\nu({\bf r})]
= \frac{3}{4} \left ( \frac{3}{\pi} \right ) ^{1/3}
e^2h_{Coul}\int d{\bf r}\rho_\pi^{4/3}({\bf r})
\left [\frac{\rho_\pi({\bf r})+\rho_\nu({\bf r})}
{\rho _0}\right]^\sigma.
\eeq
Here $\rho _0 = 0.16$ fm$^{-3}$, $\sigma = 1/3$ and from a fit of the
masses and radii of 100 medium and heavy spherical nuclei Fayans
determined $h_{Coul}=0.941$, see also Ref. \cite{bab}. The Coulomb
correlation energy determined by Fayans thus almost exactly cancels
the Coulomb exchange energy, which in the Slater approximation is
formally given by the same formula with $h_{Coul}=-1$ and $\sigma=0$,
namely
\beq
F_{Coul}^{Fock}[\rho_\pi({\bf r}),\rho_\nu({\bf r})]
\approx \frac{3}{4} \left ( \frac{3}{\pi} \right ) ^{1/3}
e^2\int d{\bf r}\rho_\pi^{4/3}({\bf r}).
\eeq
A quick estimate of the Coulomb correlation energy contribution shows that
\beq
F_{Coul}^{corr}\approx
\frac{3}{4}\left (\frac{3\rho _0}{2\pi}\right)^{1/3}
e^2h_{Coul}\int d{\bf r}\rho_\pi({\bf r})\approx
0.4\,Z{\rm{MeV}},
\eeq
which leads to a typical shift $\Delta \varepsilon_{\pi\;l} \approx
0.4 {\rm{MeV}}$ of the same order of magnitude as estimated by us
independently. An upward shift of this magnitude of the last occupied
proton level in a nucleus near the proton drip line is equivalent to a
shift of the proton drip line in the direction of decreasing $Z$ by a
few units. A shift of this magnitude for a neutron level would be
equivalent to changing the mass number by up to 5 units, see for
example \cite{bm1}. A similar shift due to the Coulomb interaction
arises for neutron levels, but we shall not discuss this mechanism
here (see however Fig. 1 and the ensuing discussion).

In a recent preprint Brown {\it et al.} \cite{brl} show that one can
generate essentially an innumerable range of phenomenological nuclear
density functionals of the Skyrme type, with which the Nolen--Shiffer
anomaly can be accounted for, while at the same time nuclear masses
and radii can be calculated selfconsistently with very good
accuracy. This lack of uniqueness for the nuclear density functional
is due to the lack of an uncontroversial theoretical underpinning. It
is hoped that the present results will bring the goal of a complete
theoretical understanding within our grasp.

One can show that there is another related effect due to the Coulomb
correlation energy, a noticeable renormalization of the proton
effective mass near the Fermi surface. In the case of homogeneous
nuclear matter the effective mass is defined as \cite{lan}
\beq
\frac{1}{m^*}=\left.\frac{1}{p_F}
\frac{d\varepsilon(p)}{dp}\right|_{p=p_F},
\eeq
where $p_F$ is the Fermi momentum. One can show that the effective
mass renormalization $\Delta m$ can be obtained from the relation
\beq
\frac{\Delta m}{m^*(m^*+\Delta m)}= \frac{e^2}{p_F}
\frac{d}{dp}\left \{ \int \frac{d{\bf q}d\omega}{(2\pi)^3}
\frac{\delta\chi_0(q,i\omega)}{\delta n (p)}
\frac{1}{q^2[1-R(q,i\omega)\chi_0(q,i\omega)]^2} \right \},
\eeq
where $m^*$ is the proton effective mass computed in the absence of
the Coulomb correlation energy and $n(p)$ is the quasiparticle
occupation number probability of the state with linear momentum
$p$. Using the following approximate formula (which becomes an
identity at the point where the compressibility is vanishing)
\beq
\left .\frac{d}{dp}\frac{\delta \chi_0(q,\omega)}{\delta
n(p)}\right|_{p=p_F}
\approx -\frac{4\pi}{p_F^2}\delta(p_F-|{\bf p+q}|)
\delta(\omega){\bf p}\cdot({\bf p+q}),
\eeq
one can reduce the above expression for the effective mass shift to
\beq
\frac{1}{m^*+\Delta m}\approx \frac{1}{m^*}
+\frac{e^2}{2\pi p_F}\int^1_{-1}
\frac{x\,dx}
{[1-R(q(x),0)\chi_0(q(x),0)]^2}, \label{meff}
\eeq
where $q(x)=p_F\sqrt{2(1-x)}.$ At the point where the compressibility
tends to zero, the denominator $[1-R(q(x),0)\chi_0(q(x),0)]$ vanishes
at $x=1\,(q=0)$. We have shown in Ref. \cite{av} that the nuclear
surface of finite nuclei is rather close to this regime. Since the
integrand in Eq. (\ref{meff}) is positive, the integral diverges and
the effective mass thus vanishes in an infinite homogeneous system.
In finite nuclei this divergence is smoothed out and the pole
singularity becomes a narrow surface peak \cite{av}. The net effect is
that the proton effective mass becomes smaller than the neutron
effective mass, and therefore the proton level density is on average
smaller than the neutron level density in nuclei. B.A. Brown argues
that a somewhat similar in medium effective nucleon mass
renormalization can account for the Nolen--Schiffer anomaly
\cite{bab}. An earlier QCD sum--rule approach \cite{hhp} substantiate
such a claim. This effect, as well as the contributions arising from
CSB forces \cite{csb}, lead mainly to volume effects, while the
many--body mechanism discussed by us leads to a surface contribution.

There is thus hope not only to generate in the near future extremely
accurate nuclear density functionals, but also to be able to
understand the nature of various rather subtle contributions. We
anticipate as well that the study of the Nolen--Schiffer anomaly
within the framework of the density functional approach will allow us
to estimate the magnitude and the character (volume versus surface) of
the CSB many--body effects.

\section*{Acknowledgments}

We thank G.F. Bertsch, E.M. Henley and G.A. Miller for discussions and
a critical reading of the manuscript.  This research was funded in
part by DOE and by INTAS under Grant No.  INTAS-OPEN-97-603. VRS would
like to thank the Department of Physics of the University of
Washington, for hospitality, where part of this work was done. We
thank also B.A. Brown for sending us the preprint \cite{brl} prior to
its publication.


\begin{thebibliography}{99}
\bibitem{erratum} In Ref. \cite[a)]{av} the lhs of the Eq. (3) should
be integrated over $\omega$.
\end{thebibliography}
\end{document}